\documentclass{article}
\usepackage{spconf,amsmath,graphicx}
\usepackage{booktabs}
\usepackage{caption}
\usepackage{hyperref}
\usepackage{amssymb}

\DeclareMathOperator*{\argmin}{arg\,min}
\usepackage{bm}
\usepackage{multicol, multirow}
\usepackage{comment}
\usepackage{tikz}
\usepackage{textcomp}
\usepackage{pifont}
\usepackage{xpatch}
\newcommand{\cmark}{\ding{51}}%
\newcommand{\xmark}{\ding{55}}%

\def\L{{\cal L}}

\title{Using Adapters to Overcome Catastrophic Forgetting in End-to-End Automatic Speech Recognition}
\name{Steven Vander Eeckt, Hugo Van hamme}
\address{KU Leuven \\ 
     Department Electrical Engineering ESAT-PSI, Leuven, Belgium\\
     \textit{\{steven.vandereeckt, hugo.vanhamme\}@esat.kuleuven.be}}
\newcommand\blfootnote[1]{%
  \begingroup
  \renewcommand\thefootnote{}\footnote{#1}%
  \addtocounter{footnote}{-1}%
  \endgroup
}

\begin{document}
\ninept
\maketitle
\begin{abstract}
\blfootnote{Research supported by Research Foundation Flanders (FWO) under grant S004923N of the SBO programme.}
Learning a set of tasks in sequence remains a challenge for artificial neural networks, which, in such scenarios, tend to suffer from Catastrophic Forgetting (CF). The same applies to End-to-End (E2E) Automatic Speech Recognition (ASR) models, even for monolingual tasks. In this paper, we aim to overcome CF for E2E ASR by inserting adapters, small architectures of few parameters which allow a general model to be fine-tuned to a specific task, into our model. We make these adapters task-specific, while regularizing the parameters of the model shared by all tasks, thus stimulating the model to fully exploit the adapters while keeping the shared parameters to work well for all tasks. 
Our method outperforms all baselines on two monolingual experiments while being more storage efficient and without requiring the storage of data from previous tasks. 
\end{abstract}

\section{Introduction}
Artificial Neural Networks (ANN) have considerably transformed Automatic Speech Recognition (ASR). With End-to-End (E2E) models, they have become ASR's state of the art. However, while ANNs are very successful at learning single tasks (e.g. learning one accent of English) or multiple tasks jointly (e.g. learning multiple accents of English simultaneously), they experience great difficulty when learning a set of tasks in sequence (e.g. learning English accents one by one). If nothing is done to prevent it, ANNs will in that case suffer from Catastrophic Forgetting (CF) \cite{catastrophicforgetting}. Measures must be taken to allow ANNs to learn tasks in sequence (without CF), i.e. to enable them to learn continually, called Continual Learning (CL). 

Within the image classification community, CL has received much attention. Consequently, many algorithms to enable CL, called CL methods, have been proposed. They can be categorized into three groups \cite{defy}: (a) regularization-based methods, such as \cite{ewc,mas},  add a regularization term to the loss to prevent forgetting; (b) rehearsal-based methods store a set of representative samples of previous tasks to rehearse during the training of new tasks - e.g. \cite{gem,er}; (c) architectural-based methods, e.g. \cite{pathnet, hat}, update the architecture of the model when new tasks arrive, thus preventing forgetting. 
Contrary to image classification, CL for ASR has received little attention. \cite{clhmmdnn,clacousticmodel} focus on CL for Hidden Markov Model based ASR models. \cite{fu2021incremental} applies \cite{lwm} to a E2E ASR model. \cite{lifelongasr, eeckt2021continual} implement, respectively, four and nine CL methods for E2E ASR. Finally, \cite{cwav2vec} focuses on semi-supervised E2E ASR, aiming to overcome CF in the wav2vec 2.0 \cite{wav2vec2} model. To do this, they insert adapters \cite{adapters}, which they make task-specific, and adapt to the new task, while freezing the shared parameters. 
Consequently, CL remains a major challenge for E2E ASR, which means that its models cannot be extended to new languages, dialects, or simply new data that becomes available, without suffering from catastrophic forgetting; unless all previous data is re-introduced during training, a strategy which soon becomes too expensive, both in terms of storage and training time. CL should enable E2E ASR models to exploit all the data at its disposal, whenever it becomes available, without suffering from CF. 

In this paper, we address this challenge for E2E ASR. To this end, we use, similar to \cite{cwav2vec}, adapters, which we insert into the model and make task-specific, meaning that each task uses its own adapters. In addition, we protect the shared parameters and thus prevent CF. We test our method on two monolingual experiments and find that it outperforms all baselines, while being more storage efficient and without requiring the storage of some past data. 

\section{Continual Learning for E2E ASR}

\subsection{E2E ASR Model}
The model is a hybrid encoder-decoder E2E model, consisting of a Conformer  \cite{conformers} or Transformer \cite{transformer} encoder and Transformer decoder. As the model is hybrid, a CTC loss is computed on the encoder's output, in addition to a decoder cross-entropy (CE) loss: their weights are $c=0.3$ and $1-c=0.7$, respectively. The output of the model is a probability distribution over $v$ word pieces. We do not use an external Language Model (LM).  
We denote $\L(X, y; \theta)$ the loss of the model with parameters $\theta \in \mathbb{R}^N$ ($N$ is the number of parameters), $X \in \mathbb{R}^{F \times f}$ the input utterance consisting of $F$ frames of dimension $f$, and $y$ the ground truth transcription of $w$ word pieces.

\subsection{Objective of Continual Learning}
\label{subsec:objective}
Given $D_1, D_2, ..., D_T$, the labeled data (containing pairs $(X, y)$) of $T$ tasks, the objective is to learn these tasks in sequence such that:
\begin{equation}
    \theta^T = \argmin_\theta \sum_{i=1}^T \sum_{(X, y) \in D_i} \L(X, y; \theta)
\end{equation}
where $\theta^T$ is adapted from $\theta^{T-1}$, the parameters of the model after training $T-1$ tasks. When learning task $T$, the model no longer has access to data from previous tasks. The objective is thus to learn the new task $T$ from $D_T$, while retaining the performance on the previous tasks $1$ to $T-1$ without having access to $D_1, ..., D_{T-1}$. Thus, the model should, on the one hand, have the ability to learn the new task well, called \textit{forward transfer} \cite{biesialska-etal-2020-continual}; on the other hand, it should be able to retain its performance on previous tasks, called \textit{knowledge retention} \cite{biesialska-etal-2020-continual}. \textit{Forward transfer} and \textit{knowledge retention} are two desiderata of the continually trained model. In addition, we add, like \cite{biesialska-etal-2020-continual}, \textit{fixed model capacity}, which demands that the storage requirements of the model do not increase with the number of tasks. 


\subsection{Inserting Adapters}
Similar to \cite{cwav2vec}, we consider adapters \cite{adapters}, small architectures inserted to fine-tune a model trained on a general task to specific tasks.  The adapters from \cite{cwav2vec} were originally used in Neural Language Processing (NLP) in a LM \cite{adapters}. Since the adapters consist of few parameters, the adaptation of the pre-trained LM to the specific NLP tasks is very easy and fast. 
In ASR, adapters have been used to train multilingual models \cite{multilingual_adapter2019, winata21_interspeech}, with each language having its own adapters. In \cite{cross_lingual_adapters}, the adapters of different languages are put in parallel and the model must learn which adapters to use.  In \cite{Gong_2021}, adapters are used to adapt an ASR model to specific accents unseen during training. In semi-supervised ASR, \cite{2022asr_ssl_adapter} uses adapters to fine-tune the pre-trained wav2vec 2.0 model to downstream tasks. 

Adapters are attractive for CL because they allow the model to exploit task-specific parameters for each task, while only slightly increasing the number of parameters, thus approximately satisfying \textit{fixed model capacity}. For that reason, we, too, consider adapters, namely the same adapters as in \cite{winata21_interspeech}, where they were used for multilingual ASR with a hybrid CTC-Transformer. These adapters consist of: (i) a layer-norm layer; (ii) a down projection to dimension $d$; (iii) a ReLU activation; (iv) an up projection to the original dimension; (v) a skip connection between the adapter's input and output. As in \cite{winata21_interspeech}, we insert adapters between the attention and feedforward sublayers of the Transformer layers. For the Conformer encoder, we insert the adapters between the convolution module and last feedforward layer. Since preliminary experiments showed that CF mainly occurs in the encoder for monolingual tasks, which, intuitively, makes sense since the tasks are part of the same language, we insert the adapters only in the encoder layers. 

For the dimension $d$ of the adapters, we consider $d=32$. With $d=32$, the number of parameters per adapter is $N_a=17k$. By introducing adapters, the model consists, after learning $T$ tasks, of parameters $\theta^T=\{\theta^{T,S}, \alpha^1, \alpha^2, ..., \alpha^T\}$, with $\theta^{T,S} \in \mathbb{R}^{N}$ the shared parameters and $\alpha^j \in \mathbb{R}^{N_a}$ the adapter parameters for task $j \leq T$. For a new task $t$, the adapter parameters $\alpha^t$ are initialized from $\alpha^{t-1}$.


\subsection{Protecting the Shared Parameters}
\label{subsec:protect}
As the number of task-specific parameters in the adapters is small, most of the model is still shared across tasks and susceptible to CF. \cite{cwav2vec} deals with this by freezing the shared parameters. We, too, consider this method, called \textit{A/Freeze}. In addition, we consider \textit{A/CFT (Cautious Fine-Tuning)} which consists of two stages: 1) train the adapters of the new task while freezing the shared parameters (as in {A/Freeze}); 2) adapt the entire model with a ten times smaller learning rate, i.e. if $lr_\text{ft}$ is the learning rate to adapt a trained model to a new task, then use $0.1lr_\text{ft}$. In the first stage, the new task's adapters are trained to work well with the 'old' shared model, thus updating the shared parameters cautiously might improve on the new task while preventing forgetting of the old tasks. 

\subsection{Exploiting the Adapters}
\label{subsubsec:exploit}
Since the adapters are task-specific, the model should, when learning a new task, exploit the adapters as much as possible. While guaranteed for subsequent tasks (as the shared parameters are either frozen or regularized, while the adapter parameters are not), this is not the case for the initial task which the model learns. Ideally, when training the model for the first time, the model should keep the shared parameters 'general', while the adapters should be used to store the 'task-specific information'. To stimulate this, we apply weight decay to the shared parameters (but not to the adapter parameters) during training of the first task. For the weight $\lambda_{wd}=10^\omega$ of the weight decay loss, we take the largest $\omega < 0$ for which the model learns the new task at most $1\%$ worse than the model without weight decay.

\subsection{Decoding without Task Label}
\label{subsubsec:task_label}
One disadvantage of the adapter-based methods is that, even at inference time, they require the task label of the utterance, to know which adapters to use. When task labels are not available, one way to infer them is to send the utterance $T$ times (if the model has seen $T$ tasks) through the model, using adapter parameters $\alpha^i$ for each task $i \leq T$. We can then consider the output (out of the $T$ outputs) with the highest likelihood. Using the hybrid CTC/decoder output, we refer to this as \textit{conf-infer}. As this might become slow as $T$ increases, we consider another alternative, \textit{avg-apt}, in which we send utterances of which the task is not known through average adapters, thus with parameters $\tilde{\alpha}=1/T\sum_{i=1}^T \alpha^{i}$.

\section{Experimental set-up}
\label{sec:exp}

\begin{table*}
    \centering
    \caption{Results after training the six CV tasks in sequence. WER for each task (evaluated on final model) and, AVG, BWT, FWT and COV are reported. 'NTL' refers to 'No task label'. If \cmark, the model does not require the task label at inference time; \xmark $\:$ indicates that it does. Methods with 'A/' in their name use adapters. Best models are in bold. }
    \begin{tabular}{l c c c c c c c c r r r}
    & & \multicolumn{6}{c}{\textit{WER per task}} & \multicolumn{4}{c}{\textit{Summary}} \\
    \cmidrule(lr){3-8} \cmidrule(lr){9-12} 
    Model & NTL & T1--{US} & T2--{ENG} & T3--{AUS} & T4--{IND} & T5--{SCO} & T6--{IRE} & \textbf{AVG}$\downarrow$ & \textbf{BWT}$\uparrow$ & \textbf{FWT}$\uparrow$ & \textbf{COV}$\uparrow$  \\
    \toprule
    Sep. Model & \xmark & 17.3 & 10.8 & 10.6 & 16.7 & 12.1 & 11.4 & 13.14 & 0.0 & 0.0 & 100.0$\%$ \\
    Sep. Enc. & \xmark & 17.3 & 11.3 & 11.3 & 17.2 & 12.4 & 10.6 & 13.34 & 0.0 & -0.2 & $90.3\%$\\
    Sep. Enc.-FF & \xmark & 17.3 &  11.6 & 11.2 & 17.5 & 13.2 & 10.8 & 13.61 & 0.0 & -0.5 & 77.7$\%$   \\
    \midrule 
    Fine-Tuning & \cmark & 19.4 & 12.7 & 14.0 & 20.6 & 13.4 & 11.4 & 15.25 & -2.5 & 0.0 & 0.0$\%$ \\
    LWF & \cmark & 19.0 & 12.2 & 13.7 & 20.7 & 12.6 & 11.1 & 14.88 & -2.3 & +0.2 & 17.5$\%$ \\
    ER & \cmark & 18.1 & 12.0 & 13.0 & 18.6 & 12.7 & 10.8 & 14.20 & -1.3 & +0.1 & 49.8$\%$ \\
    KD & \cmark & 18.5 & 12.3 & 14.0 & 19.8 & 12.9 & 10.7 & 14.67 & -1.9 & +0.1 & 27.3$\%$ \\
    \midrule
    A/Freeze & \xmark & 16.6 & 12.1 & 11.3 & 19.5 & 14.2 & 11.2 & 14.15 & 0.0 & -1.3 & 52.0$\%$ \\
    A/CFT & \xmark & 17.2 & 11.3 & 11.2 & 17.0 & 13.1 & 11.0 & \textbf{13.45} & -0.5 & 0.0 & \textbf{85.0}$\%$ \\
    \phantom{xx}+ avg-apt & \cmark & 16.3 & 11.8 & 12.6 & 17.8 & 13.4 & 10.6 & 13.75 & -0.2 & -0.6 & 71.1$\%$ \\
    \phantom{xx}+ conf-infer & \cmark & 16.5 & 11.4 & 11.3 & 17.4 & 12.6 & 10.7 & \textbf{13.31} & -0.0 & -0.1 & \textbf{91.7}$\%$ \\
    \bottomrule
    \end{tabular}
    \label{tab:cv}
\end{table*}

For all our experiments, we use the ESPnet library \cite{watanabe2018espnet}.

\textbf{Data.} For the first experiment, we consider the Common Voice (CV) 7.0 English dataset \cite{commonvoice}, which is split into six dialects learned in the following order: United States (US), England (ENG), Australia (AUS), India (IND), Scotland (SCO), Ireland (IRE). US and ENG are the largest tasks, with 350k and 116k utterances, respectively; the rest of the tasks are smaller and their order is randomized. For the second experiment, we use the Corpus Gesproken Nederlands (CGN) \cite{cgn} dataset, consisting of Dutch speech from Belgium (BE) and the Netherlands (NL). To obtain four tasks, we first split the data into NL and BE; then we split both NL and BE into 'clean' and spontaneous speech, referring to the former tasks as {NLc} and {BEc} and to the latter as {NLs} and {BEs}, for NL and BE respectively. The tasks, between 120k and 245k utterances, are learned in this order, which starts with the largest 'clean' task and maximizes the interference between tasks (by interleaving NL and BE).

\textbf{Model.} For the CV experiments, the model consists of 12 Conformer encoder layers of attention dimension 256 and feedforward dimension 2048, and 6 Transformer decoder layers of the same dimensions. With an output of 5000 word pieces, generated on US with Sentence Piece \cite{sentencepiece}, the number of parameters is 46.8M. The adapters for one task contain 204k parameters, i.e. $0.43\%$ of the model. For the CGN experiments, the model is the hybrid CTC-Transformer from \cite{hybrid_ctctransformer}, with 12 Transformer encoder and 6 Transformer decoder layers of the same dimensions as in the CV model. The output are 300 word pieces, generated on NLc. The number of parameters is 27.4M; the adapters are thus $0.75\%$ of the model. For the CV and CGN model, it thus takes, respectively, 230 and 132 tasks for the number of parameters of the model to double, which illustrates that adapters introduce task-specific parameters while still approximately satisfying \textit{fixed model capacity}. For both models, as in \cite{eeckt2021continual}, the learning rate to adapt the model to a new task ($lr_\text{ft}$) is ten times smaller than the learning rate for the initial task (US or NLc). For both experiments, $\omega=-5$ for the weight decay (see Sec. \ref{subsubsec:exploit}). 

\textbf{Baselines.} We consider following baselines: (i) Fine-Tuning: naively learns the new task, suffering from CF; (ii) Separate Model (Sep. Model): each task $t$ trains a separate model $\theta^{t}$, initialized from $\theta^{t-1}$ (Sep. Model thus stores $t$ models); (iii) Separate Encoder (Sep. Enc.): same as Sep. Model, but the decoder is shared and frozen (after the first task); (iv) Separate Encoder-Feedforward (Sep. Enc-FF): same as Sep. Enc., but all attention (and convolution) modules are also shared and frozen; (v) Learning Without Forgetting (LWF) \cite{lwf}: regularization-based method which uses the new task's data to distill knowledge from the old to new model - LWF was the best memory-free CL method from \cite{lifelongasr, eeckt2021continual}; (vi) Knowledge Distillation (KD): a rehearsal-based method and the best CL method from \cite{eeckt2021continual} - works the same as LWF but with a memory to distill knowledge from old to new model; (vii) Experience Replay (ER): a rehearsal-based method and second-best CL method from \cite{eeckt2021continual} - trains jointly on the current task's mini-batch and a mini-batch sampled from a memory. None of these methods uses adapters. We call (ii)-(iv) 'bounds' rather than 'baselines', since they violate \textit{fixed model capacity} (by introducing a large number of task-specific parameters) and are used to see what the adapter-based methods are capable of. For (iii)-(iv), we consider making (the feedforward layers of) the encoder task-specific, since preliminary experiments showed that these layers are most susceptible to forgetting. The rehearsal-based baselines have access to a memory of 2000 and 1000 uniformly sampled utterances for the CV and CGN experiments, respectively. The size of the memory is fixed, even as tasks are added. For the hyper-parameters of the baselines, we use the values of \cite{eeckt2021continual}. 

\textbf{Metrics.} To assess the models, we consider four metrics, based on WER (expressed in percentages). Average WER (AVG), the most important metric, is computed by taking the average of WERs on all seen tasks after learning the last one. Backward Tranfer (BWT) \cite{gem} measures the extent to which \textit{knowledge retention} is achieved. Negative BWT indicates forgetting and thus the average increase of WER of previous tasks compared to when they were first learned. Forward Transfer (FWT) \cite{gem} measures the extent to which \textit{forward transfer} is achieved. Positive FWT indicates average decrease of WER of new tasks, compared to Fine-Tuning. Coverage (COV) \cite{clacousticmodel} measures to which extent the models close the gap between what we consider as worst case (Fine-Tuning) and as best case (Sep. Model) in terms of AVG. 100$\%$ COV would mean that the AVG of the model equals Sep. Model's, for 0$\%$ COV it equals Fine-Tuning's.


\section{Results} 
\begin{table*}
    \centering
    \caption{Results after learning the four CGN tasks in sequence. WER for each task (evaluated on final model) and, AVG, BWT, FWT and COV are reported. 'NTL' refers to 'No task label'. If \cmark, the model does not require the task label at inference time; \xmark $\:$ indicates that it does. Methods with 'A/' in their name use adapters. Best models are in bold.}
    \begin{tabular}{l c c c c c c r r r}
    & & \multicolumn{4}{c}{\textit{WER per task}} & \multicolumn{4}{c}{\textit{Summary}} \\
    \cmidrule(lr){3-6} \cmidrule(lr){7-10} 
    Model & NTL & T1--{NLc} & T2--{VLc} & T3--{NLs} & T4--{VLs} & \textbf{AVG}$\downarrow$ & \textbf{BWT}$\uparrow$ & \textbf{FWT}$\uparrow$ & \textbf{COV}$\uparrow$  \\
    \toprule
    Sep. Model & \xmark & 23.6 & 23.5 & 45.4 & 40.5 & 33.25 & 0.0 & 0.0 & 100.0$\%$ \\
    Sep. Enc. & \xmark & 23.6 & 24.9 & 47.3 & 42.9 & 34.69 & 0.0 & -1.9 & 84.2$\%$ \\
    Sep. Enc.-FF & \xmark & 23.6 & 24.6 & 48.4 & 45.3 & 35.48 & 0.0 & -3.0 &  75.5$\%$ \\
    \midrule 
    Fine-Tuning & \cmark &  39.3 & 27.3 & 62.2 & 40.5 & 42.32 & -12.1 & 0.0 & 0.0$\%$\\
    LWF & \cmark & 36.8 & 26.9 & 59.7 & 40.4 & 40.96 & -10.2 & -0.0 & 15.0$\%$ \\
    ER & \cmark & 32.3 & 26.4 & 56.6 & 41.8 & 39.29 & -7.3 & -0.8 & 33.5$\%$ \\
    KD & \cmark & 29.0 & 25.0 & 53.1 & 41.5 & 37.12 & -4.6 & -0.5 & 57.4$\%$ \\
    \midrule
    A/Freeze & \xmark & 22.9 & 27.2 & 51.0 & 50.0 & 37.78 & 0.0 & -6.3 & 50.1$\%$ \\
    A/CFT & \xmark & 25.9 & 24.0 & 48.8 & 39.9 & \textbf{34.64} & -2.4 & +0.3 & \textbf{84.7}$\%$ \\
    \phantom{xx}+ avg-apt & \cmark & 27.1 & 24.6 & 53.1 & 40.8 & 36.38 & -3.7 & -0.7 & 65.6$\%$ \\
    \phantom{xx}+ conf-infer & \cmark & 26.7 & 24.9 & 51.2 & 40.1 & \textbf{35.72} & -3.4 & -0.1 & \textbf{72.8}$\%$ \\
    \bottomrule
    \end{tabular}
    \label{tab:cgn}
\end{table*} 
\begin{table}
    \centering
    \caption{Ablation of the effect of weight decay (Sec. \ref{subsubsec:exploit}) on A/Freeze, and of the smaller learning rate and the two stages in A/CFT (Sec. \ref{subsec:protect}). The results are the first adaptation of CGN.}
    \begin{tabular}{l c r c}
    Model & \textbf{AVG}$\downarrow$ & \textbf{BWT}$\uparrow$ & \textbf{FWT}$\uparrow$  \\
    \toprule
     A/Freeze & 25.07 & 0.0 & -3.8  \\
     \phantom{x}$\rightarrow$ no weight decay & 25.75 & 0.0 & -4.9  \\
     \midrule 
    A/CFT & 23.20 & +0.0 & -0.1 \\
    \phantom{x}$\rightarrow$ with regular learning rate & 23.87 & -2.7 & +1.3 \\
    \phantom{x}$\rightarrow$ in one stage & 24.44 & -3.2 & +0.7 \\
    \bottomrule    
    \end{tabular}
    \label{tab:ablation}
\end{table}

We discuss the results of the CV experiments, shown in Table \ref{tab:cv}. \textbf{First}, it can be seen that adapters indeed improve the model's ability to learn continually and adapt to new tasks. A/Freeze, while only training 200k parameters per task ($0.43\%$ of the model), is able to close the gap between Fine-Tuning and Sep. Model by more than $50\%$. While the gap with Sep. Enc.-FF is still clear, it is relatively small, given that Sep. Enc.-FF trains 25.2M parameters or $53.9\%$ of the model.  \textbf{Second}, while freezing the shared parameters, as in A/Freeze, results in zero forgetting (by definition), the performance can be improved by regularizing but adapting the shared parameters. Indeed, A/CFT outperforms A/Freeze, achieving a much better FWT, even if at the cost of a slightly worse BWT. It closes the gap between worst and best case by 85.0$\%$. Consequently, A/CFT outperforms all baselines by a large margin, including Sep. Enc.-FF, while having only 200k task-specific parameters, compared to the latter's 25.2M. In addition, A/CFT even comes very close to Sep. Enc.'s performance, which has 33.5M task-specific paramaters or 71.5$\%$ of the model. \textbf{Third}, with their performance, A/Freeze and A/CFT outperform all baselines, the latter even by a large margin. Nevertheless, the adapter-based methods require task labels, while the baselines do not, making the comparison for scenarios in which the task labels are unknown at inference time unfair. In Sec. \ref{subsubsec:task_label}, we proposed two ways to overcome the adapter-based methods' need for task labels. Applied to A/CFT, we find that the deterioration in performance (compared to the case where the task label is known), is small. For avg-apt, there is indeed a small deterioration, though A/CFT with avg-apt still closes the gap by more than $70\%$, outperforming all baselines. For conf-infer, there is even an improvement, with A/CFT now outperforming Sep. Enc. and closing the gap between worst and best case by $91.7\%$, almost matching the performance of Sep. Model. Nevertheless, a disadvantage of conf-infer is that the time to infer the task increases with the number of tasks. 

We proceed with the CGN experiments, of which the results are shown in Table \ref{tab:cgn}. \textbf{First}, we note that A/Freeze and A/CFT are again very effective to enable CL, both reaching a very similar COV as in the CV experiments. In addition, the gap with the baselines is similar, except that KD is now slightly better than A/Freeze. A/CFT once again outperforms all baselines by a large margin. Moreover, it even outperforms Sep. Enc.-FF, which has 12.6M task-specific parameters or 46.0$\%$ of the model, and matches the performance of Sep. Enc, which even has 17.7M task-specific parameters or 64.5$\%$ of the model. \textbf{Second}, for the case when the task labels are not known at inference time, the results remain similar to those from the CV experiments. A/CFT with avg-apt experiences a slightly larger deterioration as for CV, though still reaching a COV of 65.6$\%$ and outperforming all baselines. For conf-infer, the deterioration is smaller, reaching a COV of $72.8\%$, which is again slightly worse than for CV, where it even outperformed A/CFT with known task labels.

\textbf{Satisfying the Desiderata.} In Sec. \ref{subsec:objective}, we stated three desiderata of a CL method. Given their very small number of task-specific parameters, we concluded, in Sec. \ref{sec:exp}, that the adapter-based methods approximately achieve \textit{fixed model capacity}. From the results of the experiments, it is also clear that A/CFT, of all methods, is best able to combine \textit{knowledge retention} and \textit{forward transfer}, and thus to learn new tasks well with minimal forgetting of old tasks. 

\textbf{CL without a memory}. \cite{lifelongasr, eeckt2021continual} found that a memory of representative utterances from previous tasks was necessary to achieve high performance in CL for E2E ASR. Even when this memory was small, the gap between the rehearsal-based methods (using this memory) and the regularization-based methods (not using this memory) was large. However, we find that this need for a memory can be overcome by introducing adapters. The combination of task-specific adapters and regularizing the shared model results in much better performance, even when the task labels are not known at inference time. Note that the adapter-based methods are also more storage efficient. Storing the memory of 2000 utterances for CV and 1000 utterances for CGN requires, expessed in number of models, 1.66 and 1.54 models, respectively. In other words, the adapter-based methods can learn 382 and 204 tasks for CV and CGN, respectively, before they match the storage requirements of the rehearsal-based methods. To conclude, compared to the rehearsal-based methods, the previous state-of-the-art, the adapter-based methods 1) perform much better; 2) are much more storage efficient; 3) do not require the storage of a memory, which may not be allowed.  

\textbf{Ablation.} Table \ref{tab:ablation} shows an ablation study on the first adaptation (from T1--NLc to T2--VLc) of CGN. \textbf{First}, it illustrates the effectiveness of applying weight decay during the initial task (Sec. \ref{subsubsec:exploit}). The model with weight decay is better able to exploit the adapters and learn new tasks, achieving a higher FWT. \textbf{Second}, it illustrates the importance of A/CFT's: 1) lower learning rate during the second stage; 2) training in two stages, i.e. by first training the adapters while freezing the shared model, before adapting the entire model. Without these two aspects, A/CFT suffers from CF; while with these two aspects, A/CFT is able to learn the new task without forgetting.

\section{Conclusion}
In this paper, we considered adapters to overcome CF in (monolingual) E2E ASR. This was done by making the adapters task-specific, allowing the model to exploit the adapter parameters to learn the new tasks. Using adapters introduced very few parameters per task, yet it turned out to be very effective. While freezing the remaining shared parameters already worked decently, A/CFT, which proceeds in two stages while using a smaller learning rate in the second stage, further improves the performance greatly, outperforming all baselines on our two experiments, using two different datasets and models. On these experiments, A/CFT even outperforms or matches the performance of baselines which have between 60 and 125 times more task-specific parameters. To overcome the need for task labels at inference time, we proposed two methods -- averaging adapters and considering the confidence -- which both resulted in, at worst, a very small deterioration in performance. Overall, thus, our method A/CFT greatly improves previous state-of-the-art, while satisfying to a much higher extent the desiderata of a CL method. In addition, A/CFT is much more storage efficient, compared to previous state-of-the-art, without requiring the storage of utterances from previous tasks in a memory, which may not always be allowed or desired.

\bibliographystyle{IEEEbib}
\bibliography{main}

\begin{thebibliography}{10}

\bibitem{catastrophicforgetting}
Michael McCloskey and Neal~J. Cohen,
\newblock ``Catastrophic interference in connectionist networks: The sequential
  learning problem,''
\newblock vol.~24 of {\em Psychology of Learning and Motivation}, pp. 109--165.
  Academic Press, 1989.

\bibitem{defy}
Matthias Delange, Rahaf Aljundi, Marc Masana, Sarah Parisot, Xu~Jia, Ales
  Leonardis, Greg Slabaugh, and Tinne Tuytelaars,
\newblock ``A continual learning survey: Defying forgetting in classification
  tasks,''
\newblock {\em IEEE Transactions on Pattern Analysis and Machine Intelligence},
  p. 1–1, 2021.

\bibitem{ewc}
James Kirkpatrick et~al.,
\newblock ``Overcoming catastrophic forgetting in neural networks,''
\newblock {\em Proceedings of the National Academy of Sciences}, vol. 114, no.
  13, pp. 3521--3526, 2017.

\bibitem{mas}
Rahaf Aljundi, Francesca Babiloni, Mohamed Elhoseiny, Marcus Rohrbach, and
  Tinne Tuytelaars,
\newblock ``Memory aware synapses: Learning what (not) to forget,''
\newblock in {\em Computer Vision -- ECCV}. 2018, pp. 144--161, Springer
  International Publishing.

\bibitem{gem}
David Lopez-Paz and Marc'Aurelio Ranzato,
\newblock ``Gradient episodic memory for continual learning,''
\newblock in {\em Proceedings of the 31st International Conference on Neural
  Information Processing Systems}, 2017, NIPS'17, p. 6470–6479.

\bibitem{er}
David Rolnick, Arun Ahuja, Jonathan Schwarz, Timothy Lillicrap, and Gregory
  Wayne,
\newblock ``Experience replay for continual learning,''
\newblock in {\em Advances in Neural Information Processing Systems}. 2019,
  vol.~32, Curran Associates, Inc.

\bibitem{pathnet}
Chrisantha Fernando et~al.,
\newblock ``Pathnet: Evolution channels gradient descent in super neural
  networks.,''
\newblock {\em CoRR}, vol. abs/1701.08734, 2017.

\bibitem{hat}
Joan Serrà, Didac Suris, Marius Miron, and Alexandros Karatzoglou,
\newblock ``Overcoming catastrophic forgetting with hard attention to the
  task,''
\newblock in {\em ICML}, 2018, pp. 4555--4564.

\bibitem{clhmmdnn}
Samik Sadhu and Hynek Hermansky,
\newblock ``{Continual Learning in Automatic Speech Recognition},''
\newblock in {\em Proc. Interspeech 2020}, 2020, pp. 1246--1250.

\bibitem{clacousticmodel}
Brady Houston and Katrin Kirchhoff,
\newblock ``{Continual Learning for Multi-Dialect Acoustic Models},''
\newblock in {\em Proc. Interspeech 2020}, 2020, pp. 576--580.

\bibitem{fu2021incremental}
Li~Fu et~al.,
\newblock ``Incremental learning for end-to-end automatic speech recognition,''
\newblock in {\em 2021 IEEE Automatic Speech Recognition and Understanding
  Workshop (ASRU)}, 2021.

\bibitem{lwm}
Prithviraj Dhar, Rajat~Vikram Singh, Kuan-Chuan Peng, Ziyan Wu, and Rama
  Chellappa,
\newblock ``Learning without memorizing,''
\newblock {\em 2019 IEEE/CVF Conference on Computer Vision and Pattern
  Recognition (CVPR)}, pp. 5133--5141, 2019.

\bibitem{lifelongasr}
Heng-Jui Chang, Hung yi~Lee, and Lin shan Lee,
\newblock ``{Towards Lifelong Learning of End-to-End ASR},''
\newblock in {\em Proc. Interspeech 2021}, 2021, pp. 2551--2555.

\bibitem{eeckt2021continual}
Steven Vander~Eeckt and Hugo Van~hamme,
\newblock ``Continual learning for monolingual end-to-end automatic speech
  recognition,''
\newblock {\em Proceedings EUSIPCO 2022}, 2022.

\bibitem{cwav2vec}
Samuel Kessler, Bethan Thomas, and Salah Karout,
\newblock ``Continual-wav2vec2: an application of continual learning for
  self-supervised automatic speech recognition,''
\newblock in {\em ICML}, 2021.

\bibitem{wav2vec2}
Alexei Baevski, Yuhao Zhou, Abdelrahman Mohamed, and Michael Auli,
\newblock ``wav2vec 2.0: A framework for self-supervised learning of speech
  representations,''
\newblock in {\em Advances in Neural Information Processing Systems}, 2020,
  vol.~33.

\bibitem{adapters}
Neil Houlsby et~al.,
\newblock ``Parameter-efficient transfer learning for {NLP},''
\newblock in {\em Proceedings of the 36th International Conference on Machine
  Learning}. 2019, vol.~97 of {\em Proceedings of Machine Learning Research},
  pp. 2790--2799, PMLR.

\bibitem{conformers}
Anmol Gulati et~al.,
\newblock ``{Conformer: Convolution-augmented Transformer for Speech
  Recognition},''
\newblock in {\em Proc. Interspeech 2020}, 2020, pp. 5036--5040.

\bibitem{transformer}
Ashish Vaswani et~al.,
\newblock ``Attention is all you need,''
\newblock in {\em Advances in Neural Information Processing Systems}. 2017,
  vol.~30, Curran Associates, Inc.

\bibitem{biesialska-etal-2020-continual}
Magdalena Biesialska et~al.,
\newblock ``Continual lifelong learning in natural language processing: A
  survey,''
\newblock in {\em Proceedings of the 28th International Conference on
  Computational Linguistics}, 2020, pp. 6523--6541.

\bibitem{multilingual_adapter2019}
Anjuli Kannan, Arindrima Datta, Tara Sainath, Eugene Weinstein, Bhuvana
  Ramabhadran, Yonghui Wu, Ankur Bapna, and Zhifeng Chen,
\newblock ``Large-scale multilingual speech recognition with a streaming
  end-to-end model,''
\newblock 2019.

\bibitem{winata21_interspeech}
Genta~Indra Winata, Guangsen Wang, Caiming Xiong, and Steven Hoi,
\newblock ``{Adapt-and-Adjust: Overcoming the Long-Tail Problem of Multilingual
  Speech Recognition},''
\newblock in {\em Proc. Interspeech 2021}, 2021, pp. 2451--2455.

\bibitem{cross_lingual_adapters}
Wenxin Hou, Han Zhu, Yidong Wang, Jindong Wang, Tao Qin, Renjun Xu, and
  Takahiro Shinozaki,
\newblock ``Exploiting adapters for cross-lingual low-resource speech
  recognition,''
\newblock {\em IEEE/ACM Trans. Audio, Speech and Lang. Proc.}, vol. 30, pp.
  317–329, 2022.

\bibitem{Gong_2021}
Xun Gong, Yizhou Lu, Zhikai Zhou, and Yanmin Qian,
\newblock ``Layer-wise fast adaptation for end-to-end multi-accent speech
  recognition,''
\newblock in {\em Interspeech 2021}. aug 2021, {ISCA}.

\bibitem{2022asr_ssl_adapter}
Bethan Thomas, Samuel Kessler, and Salah Karout,
\newblock ``Efficient adapter transfer of self-supervised speech models for
  automatic speech recognition,'' 2022.

\bibitem{watanabe2018espnet}
Shinji Watanabe et~al.,
\newblock ``{ESPnet}: End-to-end speech processing toolkit,''
\newblock in {\em Proceedings of Interspeech}, 2018, pp. 2207--2211.

\bibitem{commonvoice}
R.~Ardila et~al.,
\newblock ``Common voice: A massively-multilingual speech corpus,''
\newblock in {\em Proceedings of the 12th Conference on Language Resources and
  Evaluation (LREC 2020)}, 2020, pp. 4211--4215.

\bibitem{cgn}
Nelleke Oostdijk,
\newblock ``The spoken dutch corpus: Overview and first evaluation,''
\newblock {\em Proceedings of LREC-2000}, vol. 2, 01 2000.

\bibitem{sentencepiece}
Taku Kudo and John Richardson,
\newblock ``{S}entence{P}iece: A simple and language independent subword
  tokenizer and detokenizer for neural text processing,''
\newblock in {\em Proceedings of the 2018 Conference on Empirical Methods in
  Natural Language Processing: System Demonstrations}, 2018, pp. 66--71.

\bibitem{hybrid_ctctransformer}
Shigeki Karita et~al.,
\newblock ``{Improving Transformer-Based End-to-End Speech Recognition with
  Connectionist Temporal Classification and Language Model Integration},''
\newblock in {\em Proc. Interspeech 2019}, 2019, pp. 1408--1412.

\bibitem{lwf}
Zhizhong Li and Derek Hoiem,
\newblock ``Learning without forgetting,''
\newblock in {\em Computer Vision -- ECCV 2016}. 2016, pp. 614--629, Springer
  International Publishing.

\end{thebibliography}

\end{document}